\documentclass[conference, a4paper]{IEEEtran}
\IEEEoverridecommandlockouts
\def\BibTeX{{\rm B\kern-.05em{\sc i\kern-.025em b}\kern-.08em
    T\kern-.1667em\lower.7ex\hbox{E}\kern-.125emX}}

\usepackage{amsmath,amssymb,amsfonts,graphicx,mathtools,nccmath,amsthm,float}
\usepackage{multicol,tikz,cite,array,booktabs,url}
\usepackage{esint}
\usepackage{subcaption}
\usepackage{cite}
\usepackage{textcomp}
\usepackage{xcolor}
\usepackage{algorithm}
\usepackage{algpseudocode,algorithmicx}
\usepackage[top=.75in, left= 0.625in, right = 0.625in, bottom =1in]{geometry}

\algnewcommand{\LineComment}[1]{\State \(\#\) #1}
    
\graphicspath{{figs/}}
\DeclareGraphicsExtensions{.pdf}

\newcommand{\vect}[1]{\boldsymbol{\mathbf{#1}}}

\newtheorem{Cor}{Corollary}
\newtheorem{Lem}{Lemma}
\newtheorem{Prop}{Proposition}

\DeclareMathOperator{\diag}{diag}

\DeclareMathOperator{\trace}{Tr}
\DeclarePairedDelimiter{\norm}{\lVert}{\rVert}

\algnewcommand\algorithmicinput{\textbf{Set}}
\algnewcommand\Set{\item[\algorithmicinput]}
\algnewcommand\algorithmicinitial{\textbf{Initialize}}
\algnewcommand\Initialize{\item[\algorithmicinitial]}

\let\oldReturn\Return
\renewcommand{\Return}{\State\oldReturn}

\begin{document}

\title{Spatial Secrecy Spectral Efficiency Optimization Enabled by Reconfigurable Intelligent Surfaces}
\author{\IEEEauthorblockN{Konstantinos D. Katsanos and George C. Alexandropoulos}
\IEEEauthorblockA{Department of Informatics and Telecommunications, School of Sciences,\\ National and Kapodistrian University of Athens, Panepistimiopolis Ilissia, 15784 Athens, Greece\\
emails: \{kkatsan, alexandg\}@di.uoa.gr 
}}

\maketitle

\begin{abstract}
Reconfigurable Intelligent Surfaces (RISs) constitute a strong candidate physical-layer technology for the $6$-th Generation (6G) of wireless networks, offering new design degrees of freedom for efficiently addressing demanding performance objectives. In this paper, we consider a Multiple-Input Single-Output (MISO) physical-layer security system incorporating a reflective RIS to safeguard wireless communications between a legitimate transmitter and receiver under the presence of an eavesdropper. In contrast to current studies optimizing RISs for given positions of the legitimate and eavesdropping nodes, in this paper, we focus on devising RIS-enabled secrecy for given geographical areas of potential nodes’ placement. We propose a novel secrecy metric, capturing the spatially averaged secrecy spectral efficiency, and present a joint design of the transmit digital beamforming and the RIS analog phase profile, which is realized via a combination of alternating optimization and minorization-maximization. The proposed framework bypasses the need for instantaneous knowledge of the eavesdropper’s channel or position, and targets providing an RIS-boosted secure area of legitimate communications with a single configuration of the free parameters. Our simulation results showcase significant performance gains with the proposed secrecy scheme, even for cases where the eavesdropper shares similar pathloss attenuation with the legitimate receiver.  
\end{abstract}


\section{Introduction} \label{Sec:Intro}
The definition of the key capabilities and performance indicators of the sixth Generation (6G) of wireless networks is under the way, including demanding improvements in energy and cost efficiency, network capacity and reliability, as well localization, sensing, and trustworthiness \cite{Saad_6G_2020}. Reconfigurable Intelligent Surfaces (RISs) are recently attracting remarkable research attention  as a candidate physical-layer technology, through which a vast majority of diverse communication objectives can be offered, due to the beneficial feature of metasurfaces to dynamically control over-the-air signal propagation~\cite{George_RIS_TWC2019}. An RIS is an artificial planar structure with integrated electronic circuits \cite{WavePropTCCN}, capable to manipulate, by efficient optimization, an incoming electromagnetic field in a wide variety of functionalities \cite{Marco2019}. Physical-Layer Security (PLS), which can provide further secure guarantees especially when combined with cryptographic methods in the upper communication layers, constitutes one of the emerging wireless paradigms among the various considered RIS-enabled objectives that emerge for 6G networks~\cite{xu2023reconfiguring}.

According to numerous recent PLS studies, RISs are capable to provide significant performance gains with respect to secrecy performance metrics, such as secrecy rate \cite{Conf,katsanos2022_SSE,PLS2022_counteracting} and secrecy outage probability \cite{yang2020_SOP}, or the combination of them \cite{zhang2021_PLS_Enh,li2021outage}. However, all existing works rely on the assumption that the far-field locations of the receiving nodes are fixed, while it is evident that an RIS affects an area around it, as it was demonstrated in \cite[Sec. IV.C]{PLS2022_counteracting} through exhaustive numerical evaluations. As an exception, the recent work \cite{bai2022robust} considers a secrecy rate maximization problem, under the assumption that Eve belongs to a suspicious area. Nevertheless, the problem of determining the secure area of influence of an RIS remains still an open challenge \cite{eurasip_23}.

Motivated by the above, in this paper, we focus on investigating RIS-enabled secrecy performance under the consideration that all receivers belong to given geographical areas. Inspired by \cite{Mucchi_2017_Secr_Pressure}, we propose a novel secrecy metric, termed as spatial secrecy spectral efficiency, according to which the need for instantaneous knowledge of the receivers' exact positions is bypassed. Then, we formulate a challenging optimization problem aiming at maximizing the proposed metric, which targets the joint design of the transmit digital beamforming and the RIS phase configuration profile. Our simulation results demonstrate that, even for cases where the eavesdropper shares comparable pathloss with the legitimate receiver, the gains with the proposed design are significant.

\textit{Notations:} Vectors and matrices are denoted by boldface lowercase and boldface capital letters, respectively. The transpose, conjugate, Hermitian transpose and inverse of $\mathbf{A}$ are denoted by $\mathbf{A}^T$, $\mathbf{A}^*$, $\mathbf{A}^H$, and $\mathbf{A}^{-1}$, respectively, while $\mathbf{I}_{n}$ and $\mathbf{0}_{n}$ ($n\geq2$) are the $n\times n$ identity and zeros' matrices, respectively. ${\rm Tr}(\mathbf{A})$ and $\|\vect{a}\|_2$ represent $\vect{A}$'s trace and $\vect{a}$'s Euclidean norm, respectively, while notation $\mathbf{A}\succ\vect{0}$ ($\mathbf{A}\succeq\vect{0}$) means that the square matrix $\mathbf{A}$ is Hermitian positive definite (semi-definite). $[\mathbf{A}]_{i,j}$ is the $(i,j)$-th element of $\mathbf{A}$, $[\mathbf{a}]_i$ is $\mathbf{a}$'s $i$-th element, ${\rm diag}\{\mathbf{a}\}$ denotes a square diagonal matrix with $\mathbf{a}$'s elements in its main diagonal, and $\otimes$ stands for the Kronecker product. $\mathbb{C}$ represents the complex number set, $\jmath$ is the imaginary unit, $|a|$ and $\angle a$ denote the amplitude and the phase of the complex scalar $a$, respectively, and $\Re\{a\}$ its real part. $\mathbb{E}\{\cdot\}$ is the expectation operator and $\mathbf{x}\sim\mathcal{CN}(\mathbf{a},\mathbf{A})$ indicates a complex Gaussian random vector with mean $\mathbf{a}$ and covariance matrix $\mathbf{A}$.

\section{System Model and Problem Formulation} \label{Sec:Sys_Model} 
The considered system model consists of a Base Station (BS) equipped with $N$ antenna elements wishing to communicate in the downlink direction with a legitimate single-antenna Receiver (RX). In the latter's vicinity, there exists a single-antenna Eavesdropper (Eve). We assume that the direct links between the BS and the two receiving nodes are blocked. To overcome this situation, an RIS with $L$ unit cells is deployed near the receivers' area. We assume throughout this paper that partial Channel State Information (CSI) for both wireless links is available at the BS side \cite{PLS2022_counteracting}: the statistics of the channels $\vect{h},\,\vect{g}\in\mathbb{C}^{L \times 1}$ (for RX and Eve, respectively) are known up to their second order. On the other hand, perfect CSI is available for the BS-RIS channel $\vect{H}\in\mathbb{C}^{L \times N}$.

\subsection{Received Signal and Channel Models}\label{Sec:Signals}
To secure the confidentiality of the legitimate link, the BS applies the linear precoding vector $\vect{v} \in \mathbb{C}^{N\times 1}$ and jointly designs the RIS reflection vector (passive beamforming) $\vect{\phi}\triangleq[e^{\jmath \theta_1}\,\,e^{\jmath \theta_2}\,\,\cdots\,\,e^{\jmath \theta_L} ]^T\in\mathbb{C}^{L \times 1}$, where $\theta_{\ell}$ ($\ell = 1,2,\ldots,L$) denotes the effective phase shifting value at each $\ell$-th RIS unit element. The baseband received signals $y_{\rm RX}$ and $y_{\rm E}$ at the RX and Eve can be mathematically expressed as follows:
\begin{equation} \label{eq:y_i}
    y_i \triangleq \tilde{\vect{f}}^H \vect{v} s + n_i, 
\end{equation}
where $s \sim \mathcal{CN}(0,1)$ denotes the legitimate information symbol (in practice, it is chozen from a discrete modulation set) and $\tilde{\vect{f}}^H \triangleq \vect{f}_i^H \vect{\Phi} \vect{H}$ ($i={\rm RX}$ and ${\rm E}$) represents the end-to-end channel between the RX/Eve and BS via the RIS with $\vect{\Phi}\triangleq\diag\{\vect{\phi}\}\in\mathbb{C}^{L \times L}$. When $i={\rm RX}$, $\vect{f}_i=\vect{h}$ implying the RIS-parametrized RX-BS channel, while $i={\rm E}$ indicates the respective Eve-BS channel $\vect{f}_i=\vect{g}$. Finally, $n_i\sim \mathcal{CN}(0,\sigma^2)$ $\forall$$i$ stands for the Additive White Gaussian Noise (AWGN). It follows from \eqref{eq:y_i} that the rate for each link is given by:
\begin{equation} \label{eq:rate_i}
    \mathcal{R}_i(\vect{v},\vect{\phi}) \triangleq \log_2\left( 1 + \frac{\lvert \tilde{\vect{f}}^H \vect{v} \rvert^2}{\sigma^2} \right).
\end{equation}

We further assume that the channels $\vect{h}$ and $\vect{g}$ undergo Rician fading with Rician factor $K_r$; we study the difficult secrecy case where RX and Eve are placed such that they share similar pathloss properties. Considering a 3D coordinate system for the nodes' placement (i.e., BS, RIS, RX, and Eve), and taking into account that both the BS and RIS possess Uniform Planar Arrays (UPAs), the channel between the RX/Eve and RIS can be expressed as follows:
\begin{equation} \label{eq:channel_f}
    \vect{f} = \sqrt{\kappa_i} \left( \sqrt{\frac{K_r}{K_r+1}} \vect{f}^{\rm LOS} + \sqrt{\frac{1}{K_r+1}} \vect{f}^{\rm NLOS} \right),
\end{equation}
where $\kappa_i \triangleq \operatorname{PL}_0 \left(\frac{D_{\rm RIS,i}}{d_0}\right)^{-\alpha_{{\rm RIS},i}}$ with $\alpha_{{\rm RIS},i}$ denoting the pathloss exponent of the $i$th link, $\operatorname{PL}_0$ is the pathloss at the reference distance $d_0 = 1$m, and $D_{{\rm RIS},i} = \| \vect{p}_{\rm RIS} - \vect{p}_i \|$ is the distance between the RIS and the reception node $i$ with $\vect{p}_{\rm RIS} \triangleq [x_{\rm RIS}, y_{\rm RIS}, z_{\rm RIS}]^T$, $\vect{p}_{\rm RX} \triangleq [x_{\rm RX}, y_{\rm RX}, z_{\rm RX}]^T$ ($i=\rm RX$), and $\vect{p}_{\rm E} \triangleq [x_{\rm E}, y_{\rm E}, z_{\rm E}]^T$ ($i=\rm E$) indicating the corresponding position vectors. The Line-Of-Sight (LOS) vector $\vect{f}^{\rm LOS}$ in \eqref{eq:channel_f} is the steering vector at the RIS, which is obtained as $\vect{f}^{\rm LOS} \triangleq \vect{a}_{L_v}(\theta_{\vect{f}},\varphi_{\vect{f}}) \otimes \vect{a}_{L_h}(\theta_{\vect{f}},\varphi_{\vect{f}})$ for the case of an RIS with $L_v$ vertical and $L_h$ horizontal elements, such that $L = L_v \times L_h$, where  $\theta_{\vect{f}}$ and $\varphi_{\vect{f}}$ denote the elevation and azimuth angles of departure (AoD) with respect to the RIS, respectively. Assuming that the RIS is deployed parallel to the $yz$ plane, $\theta_{\vect{f}} = \arccos(\frac{x_{\rm RIS} - x_i}{D_{{\rm RIS},i}})$ and $\varphi_{\vect{f}} = \arctan(\frac{y_{\rm RIS} - y_i}{z_{\rm RIS} - z_i})$. For this case, the $\ell$-th entry of the vector $\vect{f}^{\rm LOS}$ can be obtained as $[\vect{f}^{\rm LOS}]_{\ell} = \frac{1}{\sqrt{L}} e^{\jmath 2\pi \frac{d}{\lambda_c} \xi_{\ell}}$ with $\xi_{\ell}$ given by:
\begin{equation} \label{eq:xi_ell}
\begin{aligned}
    \xi_{\ell} \triangleq \Bigg( &\left\lfloor \frac{\ell-1}{L_h} \right\rfloor \sin\theta_{\vect{f}}\cos\varphi_{\vect{f}}\\
    &{}+ \left( \ell-\left\lfloor \frac{\ell-1}{L_h} \right\rfloor L_h - 1  \right)\sin\theta_{\vect{f}}\sin\varphi_{\vect{f}} \Bigg),
\end{aligned}    
\end{equation}
with $d$ and $\lambda_c$ being the RIS's element spacing and the wavelength, respectively. Finally, in \eqref{eq:channel_f}, the Non-LOS (NLOS) channel component is modeled as $[\vect{f}^{\rm NLOS}]_{\ell} \sim \mathcal{CN}(0,1)$. 

\subsection{Problem Formulation}\label{Sec:Prob_Form}
Consider that RX and Eve are both located within a given geographical area $\mathcal{S} \triangleq \mathcal{S}_{\rm RX} \cup \mathcal{S}_{\rm E}$ with $\mathcal{S}_{\rm RX} \cap \mathcal{S}_{\rm E}=\emptyset$, where $\mathcal{S}_{\rm RX}$ denotes the area where the RX can be located and $\mathcal{S}_{\rm E}$ indicates the respective area for Eve. One example where this assumption of disjoint adjacent areas can be realized is the placement of the RX and Eve in two adjacent rooms in indoor environments, with the latter node targeting to eavesdrop legitimate communications. The extension to overlapping areas $\mathcal{S}_{\rm RX}$ and $\mathcal{S}_{\rm E}$ will be considered in future work. We further assume that each point of the area $\mathcal{S}$ is associated with a Probability Density Function (PDF) $u(\mathcal{S})$, which equals the joint probability of presence for RX and Eve in locations within $\mathcal{S}_{\rm RX}$ and $\mathcal{S}_{\rm E}$, respectively. Aiming at securing communications for each placement of RX within $\mathcal{S}_{\rm RX}$ and Eve inside $\mathcal{S}_{\rm E}$, we define the spatial secrecy spectral efficiency as the following spatially averaged metric: 
\begin{equation} \label{eq:Spatial_Secr_Def}
    \mathcal{R}_{\mathcal{S}} \triangleq \mathbb{E}_{\vect{h},\vect{g}}\left\{\int_{\mathcal{S}_{\rm RX}}\!\int_{\mathcal{S}_{\rm E}} \!u(\mathcal{X},\mathcal{Y}) \left[\mathcal{R}_{\rm RX} - \mathcal{R}_{\rm E}\right]^+ \!d\mathcal{X}d\mathcal{Y} \right\}\!.
\end{equation}
We next capitalize on the assumption of disjoint $\mathcal{S}_{\rm RX}$ and $\mathcal{S}_{\rm E}$ (which are assumed as 2D geographical areas for simplicity) to simplify $\mathcal{R}_{\mathcal{S}}$'s expression for its upcoming optimization.
\begin{Prop} \label{prop:Lemma_Secr_Press}
For disjoint sets $\mathcal{S}_{\rm RX}$ and $\mathcal{S}_{\rm E}$ and independent events $\mathbf{p}_{\rm RX} \in \mathcal{S}_{\rm RX}$ and $\mathbf{p}_{\rm E} \in \mathcal{S}_{\rm E}$, where for each $\mathbf{p}_{\rm RX}$ and $\mathbf{p}_{\rm E}$ holds $z_{\rm RX}=z_{\rm E}=z$, 
\eqref{eq:Spatial_Secr_Def} is re-written as:
    \begin{equation} \label{eq:Secr_Pressure}
        \mathcal{R}_{\mathcal{S}} = \mathcal{R}_{\mathcal{S}_{\rm RX}} - \mathcal{R}_{\mathcal{S}_{\rm E}},
    \end{equation}
    where $\mathcal{R}_{\mathcal{S}_{\rm RX}} \triangleq \mathbb{E}_{\vect{h}}\left\{\int_{\mathcal{S}_{\rm RX}} u_{\rm RX}(x_{\rm RX}, y_{\rm RX}) \mathcal{R}_{\rm RX} dx_{\rm RX}dy_{\rm RX}\right\}$ and $\mathcal{R}_{\mathcal{S}_{E}} \triangleq \mathbb{E}_{\vect{g}}\left\{\int_{\mathcal{S}_{\rm E}} u_{\rm E}(x_{\rm E}, y_{\rm E}) \mathcal{R}_{\rm E}dx_{\rm E}dy_{\rm E}\right\}$ 
    with notations $u_{\rm RX}(\cdot,\cdot)$ and $u_{\rm E}(\cdot,\cdot)$ representing the 2D spatial PDFs of RX and Eve, respectively.
\end{Prop}
\begin{IEEEproof}
From the previous assumptions, it can be deduced $u(\mathcal{S}) = u_{\rm RX}(\mathcal{S}_{\rm RX}) u_{\rm E}(\mathcal{S}_{\rm E})$. 
It also holds that $\int_{\mathcal{S}_{\rm RX}} u_{\rm RX}(\mathcal{X}) d\mathcal{X} = 1$ and $\int_{\mathcal{S}_{\rm E}} u_{\rm RX}(\mathcal{Y}) d\mathcal{Y} = 0$.
Since the integrand function is non-negative, Tonelli's theorem \cite{royden1988real} can be applied, which completes the proof.
\end{IEEEproof}

It can be seen from \eqref{eq:Secr_Pressure} that $\mathcal{R}_{\mathcal{S}}$ depends on $\vect{v}$ and $\vect{\phi}$. We now focus on the following design optimization problem:
\begin{align*}
\begin{split}
    \mathcal{OP}: \,\, &\max_{\vect{v}, \vect{\phi}} \quad \mathcal{R}_{\mathcal{S}}(\vect{v},\vect{\phi}) \\
    &\,\, \text{s.t.} \quad\,\,\,\,\norm{\vect{v}}_2^2 \leq P_T, \\
    &\,\, \quad\quad\quad \lvert \phi_{\ell} \rvert = 1 \,\, \forall \ell = 1,2,\dots,L,
\end{split}
\end{align*}
where $P_T$ in the first constraint is the BS's transmit power whose value should not be exceeded during any transmission slot, and the second constraint guarantees the unit-modulus property of each RIS's element. $\mathcal{OP}$ is a challenging non-convex optimization problem due to the coupling variables $\vect{v}$ and $\vect{\phi}$ and the double integration operations over the considered geographical areas $\mathcal{S}_{\rm RX}$ and $\mathcal{S}_{\rm E}$. 

\section{Spatial Secrecy Efficiency Optimization} \label{Sec:Prob_Solution}
We first approximate the objective function $\mathcal{R}_{\mathcal{S}}$, as follows. We assume, without loss of generality, that the presence of both RX's and Eve's in $\mathcal{S}_{\rm RX}$ and $\mathcal{S}_{\rm E}$, respectively, follows a uniform distribution in their respective areas. Based on this assumption, the resulting 2D spatial PDFs become $u_{\rm RX}(x_{\rm RX},y_{\rm RX}) = \frac{1}{\mathcal{S}_{\rm RX}}$ and $u_{\rm E}(x_{\rm E},y_{\rm E}) = \frac{1}{\mathcal{S}_{\rm E}}$. Hence, $\mathcal{OP}$'s objective function can be simplified to\footnote{It is noted that the operation $[\cdot]^+$ is henceforth omitted since it does not affect the derivation of $\vect{v}$ and $\vect{\phi}$. This deduces from the fact that our design target is to provide positive secrecy rates.}:
\begin{equation} \label{eq:simplified_f_sec}
    \begin{aligned}
        \mathcal{R}_{\mathcal{S}}(\vect{v},\vect{\phi}) = &{}\frac{1}{\mathcal{S}_{\rm RX}}\int_{\mathcal{S}_{\rm RX}} \mathbb{E}_{\vect{h}}\left\{\mathcal{R}_{\rm RX}\right\}dx_{\rm RX}dy_{\rm RX}\\
    &- \frac{1}{\mathcal{S}_{\rm E}} \int_{\mathcal{S}_{\rm E}} \mathbb{E}_{\vect{g}}\left\{\mathcal{R}_{\rm E}\right\}dx_{\rm E}dy_{\rm E},
    \end{aligned}    
\end{equation}
where the linearity of the integration and expectation operations was used. The considered spatial secrecy spectral efficiency can be further simplified via the following corollary. 
\begin{Cor} \label{thm:approx_pressure}
First, consider the following definitions:
\begin{equation} \label{eq:mean_covariance_f}
    \bar{\vect{f}} \triangleq \sqrt{\kappa_i}\sqrt{\frac{K_r}{1+K_r}}\vect{f}^{\rm LOS},\,\,
    \vect{M}_{\vect{f}} \triangleq \frac{\kappa_i}{1 + K_r}\vect{I}_L.
\end{equation}
Moreover, $\bar{\vect{f}}$ is the mean of channel $\vect{f}$, according to the adopted Rician channel model (based on \eqref{eq:channel_f}), and $\vect{M}_{\vect{f}}$ is its covariance matrix (with $\vect{M}_{\vect{f}}$ $\succ \vect{0}$), i.e., $\vect{f} \sim \mathcal{CN}(\bar{\vect{f}},\vect{M}_{\vect{f}})$. Let, also, the matrix $\vect{J}_i \in \mathbb{C}^{L,L}$ be defined as follows:
\begin{equation} \label{eq:matrices_J}
    \vect{J}_i \triangleq \int_{\mathcal{S}_i} \left( \vect{M}_{\vect{f}} + \bar{\vect{f}}\bar{\vect{f}}^H \right)dx_idy_i.
\end{equation}
Then, the average spatial secrecy spectral efficiency, can be approximated (having the same lower and upper bounds) as:
\begin{equation} \label{eq:f_sec_approx}
    \mathcal{R}_{\mathcal{S}}\approx \tilde{\mathcal{R}}_{\mathcal{S}}(\vect{v},\vect{\phi}) \triangleq \log_2\left( \frac{1 + \frac{\vect{v}^H\vect{H}^H\vect{\Phi}^H\vect{J}_{\rm RX}\vect{\Phi}\vect{H}\vect{v}}{\sigma^2\mathcal{S}_{\rm RX}}}{1 + \frac{\vect{v}^H\vect{H}^H\vect{\Phi}^H\vect{J}_{\rm E}\vect{\Phi}\vect{H}\vect{v}}{\sigma^2\mathcal{S}_{\rm E}}} \right).
\end{equation}
\end{Cor}
\begin{IEEEproof}
Expression \eqref{eq:f_sec_approx} can be derived using straightforward algebraic manipulations and~\cite[Lemma 1]{wang2018_multiple}. The detailed steps are omitted for brevity.
\end{IEEEproof}

The maximization of \eqref{eq:f_sec_approx} with respect to $\vect{v}$ and $\vect{\phi}$ remains a non-convex optimization problem due to the coupling of the optimization variables and the non-convex unit-modulus constraints. However, it can be efficiently solved by adopting Alternating Optimization (AO), i.e.: iterative optimizations keeping one variable fixed to obtain an expression for the other one, until a convergence criterion is satisfied, as described in the following subsections.

\subsection{Optimization over the Active BS Beamformer $\vect{v}$} \label{sec:opt_v}
The maximization problem with respect to $\vect{v}$ is a fractional QCQP problem with a convex constraint, which can be optimally solved according to the following lemma.
\begin{Lem}
The optimal expression for the linear precoding vector $\vect{v}$ is given by:
\begin{equation} \label{eq:opt_v}
    \vect{v}^{\rm opt} = \sqrt{P_T}\vect{u}_{\max}\left(\vect{P}_{\rm E}^{-1} \vect{P}_{\rm RX}\right),
\end{equation}
where $\vect{P}_i \triangleq \vect{I}_N + \frac{P_T}{\sigma^2 \mathcal{S}_i}\vect{H}^H\vect{\Phi}^H \vect{J}_i \vect{\Phi} \vect{H}$, with $i \in \{\rm RX, E\}$, and $\vect{u}_{\max}(\vect{A})$ denotes the eigenvector corresponding to the largest eigenvalue of matrix $\vect{A}$. 
\end{Lem}
\begin{IEEEproof}
We first note that the logarithmic operation can be omitted because $\log_2(\cdot)$ is a non-decreasing function. Then, it can be readily verified that the optimal $\vect{v}$ satisfies the constraint $\|\vect{v}\|^2 \leq P_T$ with equality. Based on these two key properties, the optimization problem reduces to a generalized eigenvalue problem whose optimal solution is given by \eqref{eq:opt_v}. 
\end{IEEEproof}

\subsection{Optimization over the Passive RIS Beamformer $\vect{\phi}$} \label{seq:opt_phi}
Under the assumption that $\vect{v}$ is fixed, it can be shown after some algebraic manipulations that $\tilde{\mathcal{R}}_{\mathcal{S}}(\cdot)$ reduces to:
\begin{equation}
   \tilde{\mathcal{R}}_{\mathcal{S}}(\vect{\phi}) = \frac{\vect{\phi}^H\vect{Q}_{\rm RX} \vect{\phi}}{\vect{\phi}^H\vect{Q}_{\rm E}\vect{\phi}},
\end{equation}
with $\vect{Q}_i \triangleq \frac{1}{L}\vect{I}_L + \frac{1}{\sigma^2\mathcal{S}_i}\diag\{{\tilde{\vect{v}}}\}^H\vect{J}_i\diag\{\tilde{\vect{v}}\}$, for $i \in \{\rm RX, E\}$, where $\tilde{\vect{v}} \triangleq \vect{H} \vect{v}$ and the property $\vect{\phi}^H\vect{\phi} = L$ has been invoked. Then, based on the Majorization-Minimization (MM) technique \cite{sun2016majorization}, it can be shown that the objective function can be minorized around a feasible point $\vect{\phi}_t$ by:
\begin{align}
    g(\vect{\phi}\vert\vect{\phi}_t) = &{}2\frac{\Re\{\vect{\phi}_t^H\vect{Q}_{\rm RX} \vect{\phi}_t\}}{\vect{\phi}_t^H\vect{Q}_{\rm E} \vect{\phi}_t} - \frac{\vect{\phi}_t^H\vect{Q}_{\rm RX}\vect{\phi}_t}{(\vect{\phi}_t^H \vect{Q}_{\rm E} \vect{\phi}_t )^2}\Big\{ \vect{\phi}^H \trace(\vect{Q}_{\rm E}) \vect{\phi} \notag \\
    &+ 2\Re\{ \vect{\phi}_t^H \left[ \vect{Q}_{\rm E} - \trace(\vect{Q}_{\rm E})\vect{I}_L \right] \vect{\phi} \} \Big\},
\end{align}
where $\trace(\vect{Q}_{\rm E}) > 0$ since $\vect{Q}_{\rm E} \succ \vect{0}$. Then, at the inner iteration $(t+1)$ of the MM-based algorithm, it can be shown that $\vect{\phi}_{t+1} = \exp\left\{\jmath \angle \tilde{\vect{\phi}} \right\}$, where $\tilde{\vect{\phi}}$ is given by:
\begin{equation}
    \tilde{\vect{\phi}} = \frac{\vect{Q}_{\rm RX}\vect{\phi}_t}{\vect{\phi}_t^H\vect{Q}_{\rm E}\vect{\phi}_t} - \frac{\vect{\phi}_t^H\vect{Q}_{\rm RX}\vect{\phi}_t}{(\vect{\phi}_t^H \vect{Q}_{\rm E} \vect{\phi}_t )^2}\left(\vect{Q}_{\rm E} - \trace(\vect{Q}_{\rm E})\vect{I}_L\right)\vect{\phi}_t.
\end{equation}

\section{Numerical Results and Discussion} \label{Sec:Numerical} 
In this section, we present our performance evaluation results for the spatial secrecy spectral efficiency metric for the considered RIS-enabled MISO communication system.
\begin{figure}[!t]
\centering
\includegraphics[scale=0.55]{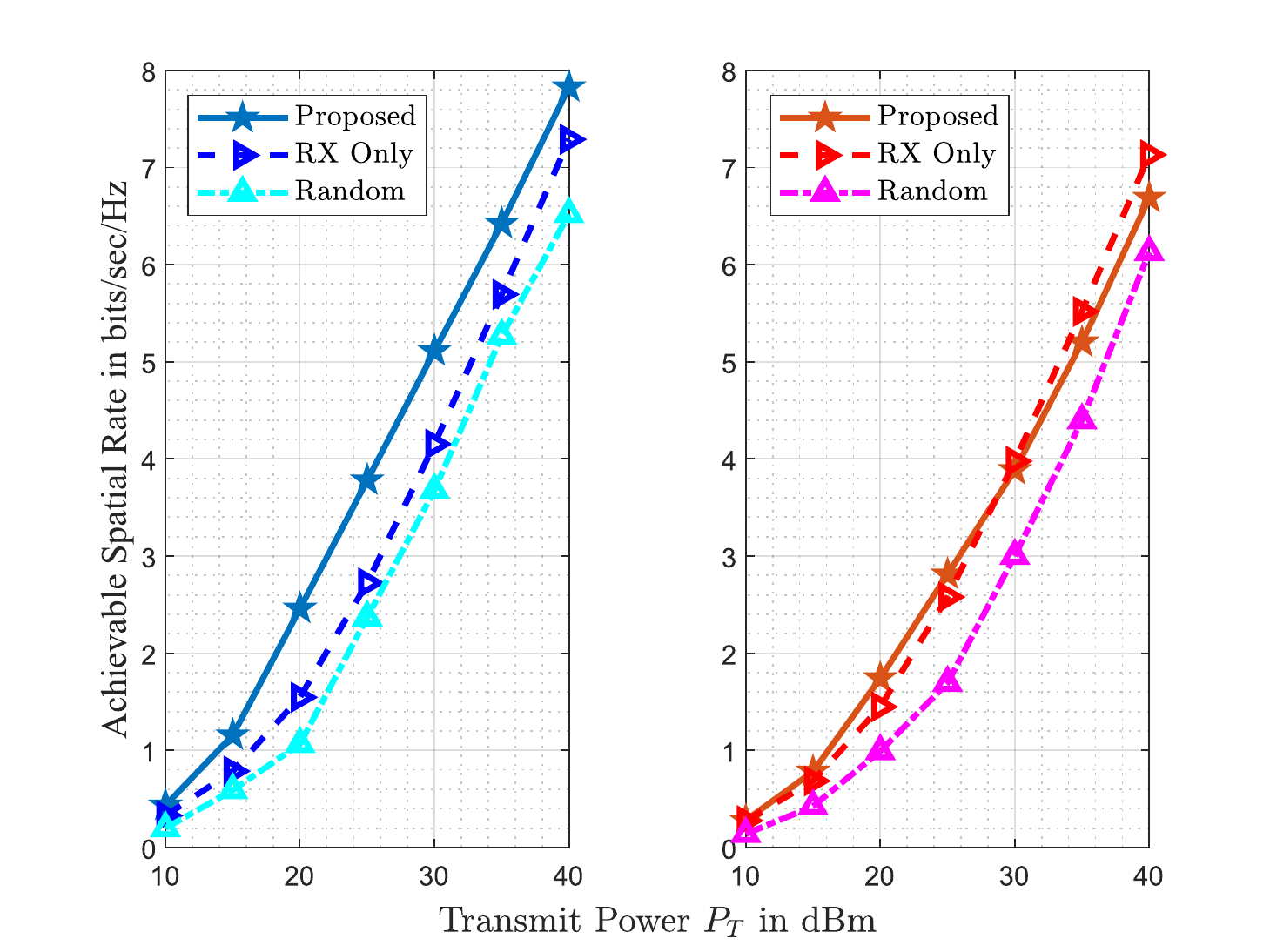}
\caption{Maximum achievable spatial rates in bits/sec/Hz at the legitimate RX (left) and the eavesdropper Eve (right) versus the transmit power $P_T$ in dBm, considering a BS with $N = 16$ antenna elements and an RIS with $L=150$ unit elements.}
\label{fig:Rates_vs_P_T}
\end{figure}


\subsection{Simulation Setup} \label{sec:Sims_Setup}
We have evaluated the achievable rates of the legitimate and eavesdropping links using expression \eqref{eq:rate_i} for RX and Eve. We have used the proposed scheme in Section \ref{Sec:Prob_Solution} for the design of the linear precoding vector and the passive beamforming vector for the RIS configuration. In our simulations, all nodes were considered placed on a $3$D coordinate system, whose exact locations are represented by the triad $(x,y,z)$ according to the orthogonal Cartesian coordinate system. We have assumed that the BS is located at the point $(0,0,7.5)\,m$, while the RIS at the position $(0,50,3)\,m$. As studied throughout the paper, both receivers are not restricted to fixed positions, but they can be uniformly located inside their respective $\mathcal{S}_i$ areas. Specifically, we have considered that $\mathcal{S}_{\rm RX}$ is centered around the point $(-15,30)\,m$ on the $xy$ plane, having a width of $24\,m$ and length $15\,m$. Similarly, the center of Eve's area was fixed at the position $(15,27.5)\,m$ sharing the same width and length as $\mathcal{S}_{\rm RX}$. This consideration results in areas sharing similar pathloss attenuation properties, serving as a worst-case scenario. It was also assumed that $z = 1.5\,m$. The channels $\vect{h}$ and $\vect{g}$ were modeled according to \eqref{eq:channel_f} with $\operatorname{PL}_0 = -30$ dB, and the pathloss exponents were set as $\alpha_{\rm RIS,i}=2.2$. For the channel matrix $\vect{H}$ between the BS and RIS, we used the Rician fading model with the same pathloss properties. In addition, we have used the following setting for the system parameters: $N = 16$, $L = 150$, $K_r = 13.2$, $\sigma^2 = -105$ dBm, and the optimization algorithmic threshold $\epsilon$ for the convergence criterion was considered equal to $10^{-5}$. For all performance evaluation results, we have used $100$ independent Monte Carlo realizations. Moreover, the matrices $\vect{J}_i$'a defined in \eqref{eq:matrices_J} were computed via numerical integration over the $\mathcal{S}_i$ areas. For comparison purposes, we have implemented two baseline schemes: \textit{i}) the ``Random'' scheme including random linear precoding and passive beamforming vectors; and \textit{ii}) the ``RX Only'' scheme according to which we maximize the spatial spectral efficiency only for RX (defined analogously to $\mathcal{R}_{\mathcal{S}}$). 

\subsection{Rate Performance vs. the Transmit Power} \label{sec:Num_over_P_T}
We first investigate the achievable spatial secrecy spectral efficiency as a function of the transmit power $P_T$. In particular, we have evaluated the achievable rates for RX and Eve for every position in the considered $\mathcal{S}_i$ areas, and illustrate their maximum values for all schemes in Fig.~\ref{fig:Rates_vs_P_T}. It can be observed that, as expected, the rate exhibits a non-decreasing trend as the transmit power increases for both links, while the values achieved for the legitimate link are higher than those for the eavesdropping link. Moreover, it is shown that the gains in the legitimate RX's rate, with respect to Eve's achievable values, are larger for the case where the proposed metric and optimized design are adopted, in contrast to the other benchmark schemes. For instance, in the low SNR regime, this difference is small, whereas for higher SNR values (specifically, $P_T \geq 35$ dBm), it reaches its maximum difference of about $1.5$ bps/Hz. This behavior can be justified by the fact that the proposed metric is dedicated on maximizing the difference between RX's and Eve's rate, while the other two schemes do not emphasize on improving their deviation. It needs to be commented, though, that the aforementioned difference between RX's and Eve's rate is quite small. However, for the considered simulation setup, the areas $\mathcal{S}_{\rm RX}$ and $\mathcal{S}_{\rm E}$ yield comparable pathloss-based properties between the two receivers, which are encapsulated into the $\vect{J}_i$ matrices, where all distance-dependent characteristics are integrated on average. Nevertheless, even for such cases where there do not exist direct links from the BS to both receiving nodes, the optimized RIS becomes capable to provide larger rates for the legitimate RX.

\begin{figure*}[!t]
\centering
    \begin{subfigure}[t]{0.43\textwidth}
    \centering
    \includegraphics[width=\textwidth]{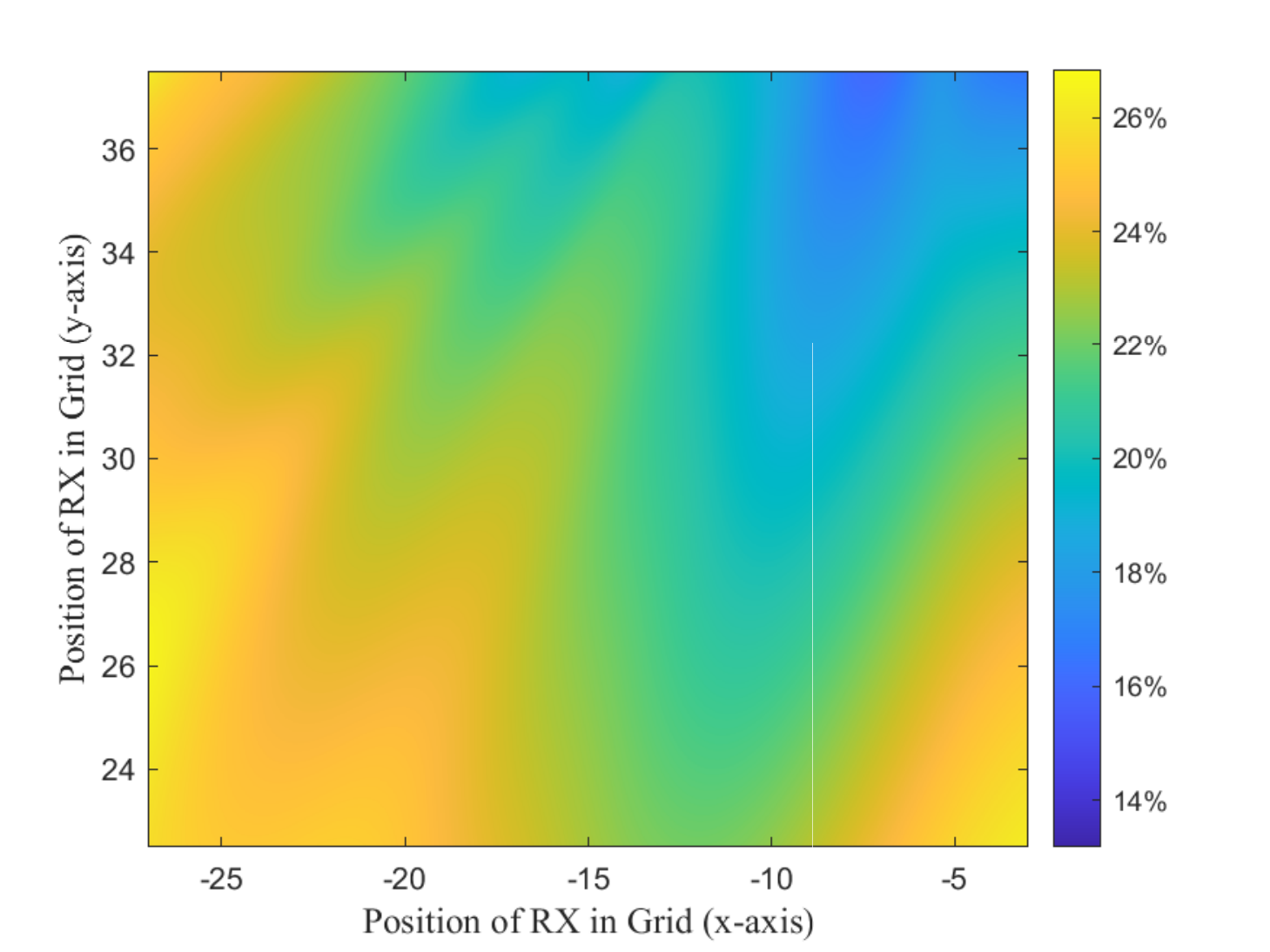}
    \caption{Spatial rate gain at RX between the random configuration and the ``RX Only'' scheme.} 
    \label{fig:Heat_RX_Gain_RX_ONLY}
\end{subfigure}\quad
\begin{subfigure}[t]{0.43\textwidth}
    \centering
    \includegraphics[width=\textwidth]{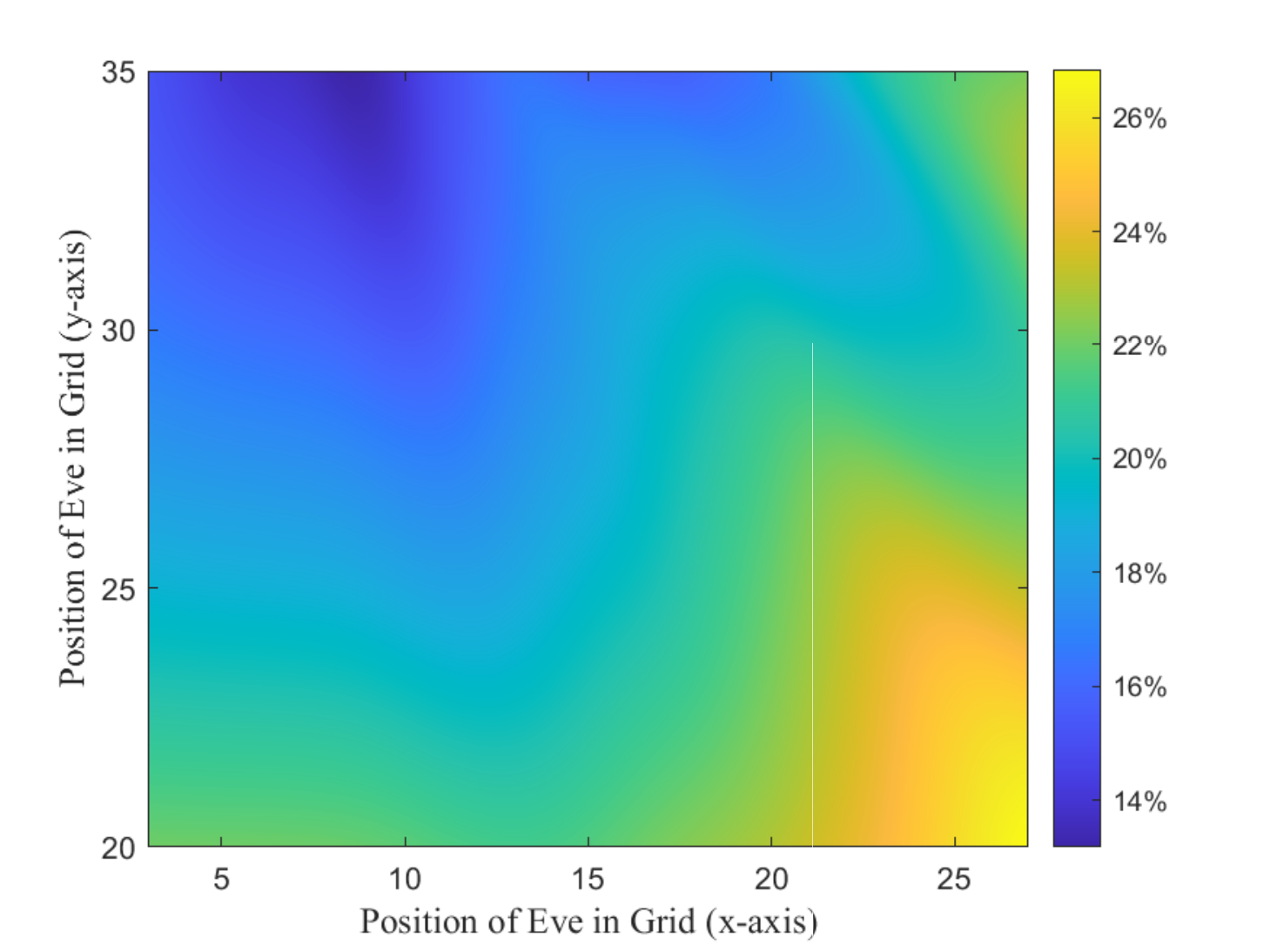}
    \caption{Spatial rate gain at Eve between the random configuration and the ``RX Only'' scheme.} 
    \label{fig:Heat_E_Gain_RX_ONLY}
\end{subfigure}\quad
 \begin{subfigure}[t]{0.43\textwidth}
    \centering
    \includegraphics[width=\textwidth]{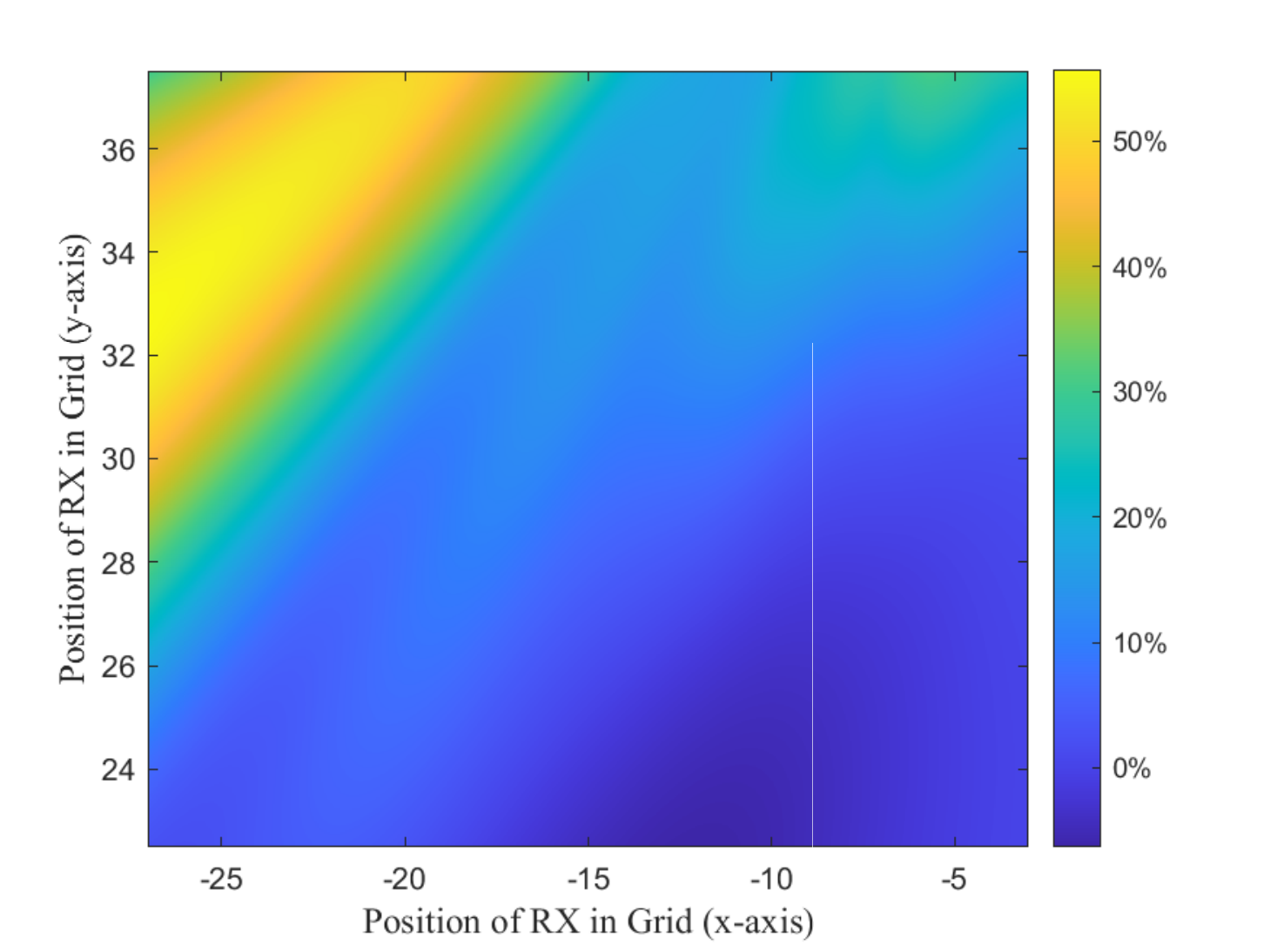}
    \caption{Spatial rate gain at RX between the random configuration and the proposed secrecy spatial rate maximization scheme.} 
    \label{fig:Heat_RX_Gain_Full}
\end{subfigure}\quad
 \begin{subfigure}[t]{0.43\textwidth}
    \centering
    \includegraphics[width=\textwidth]{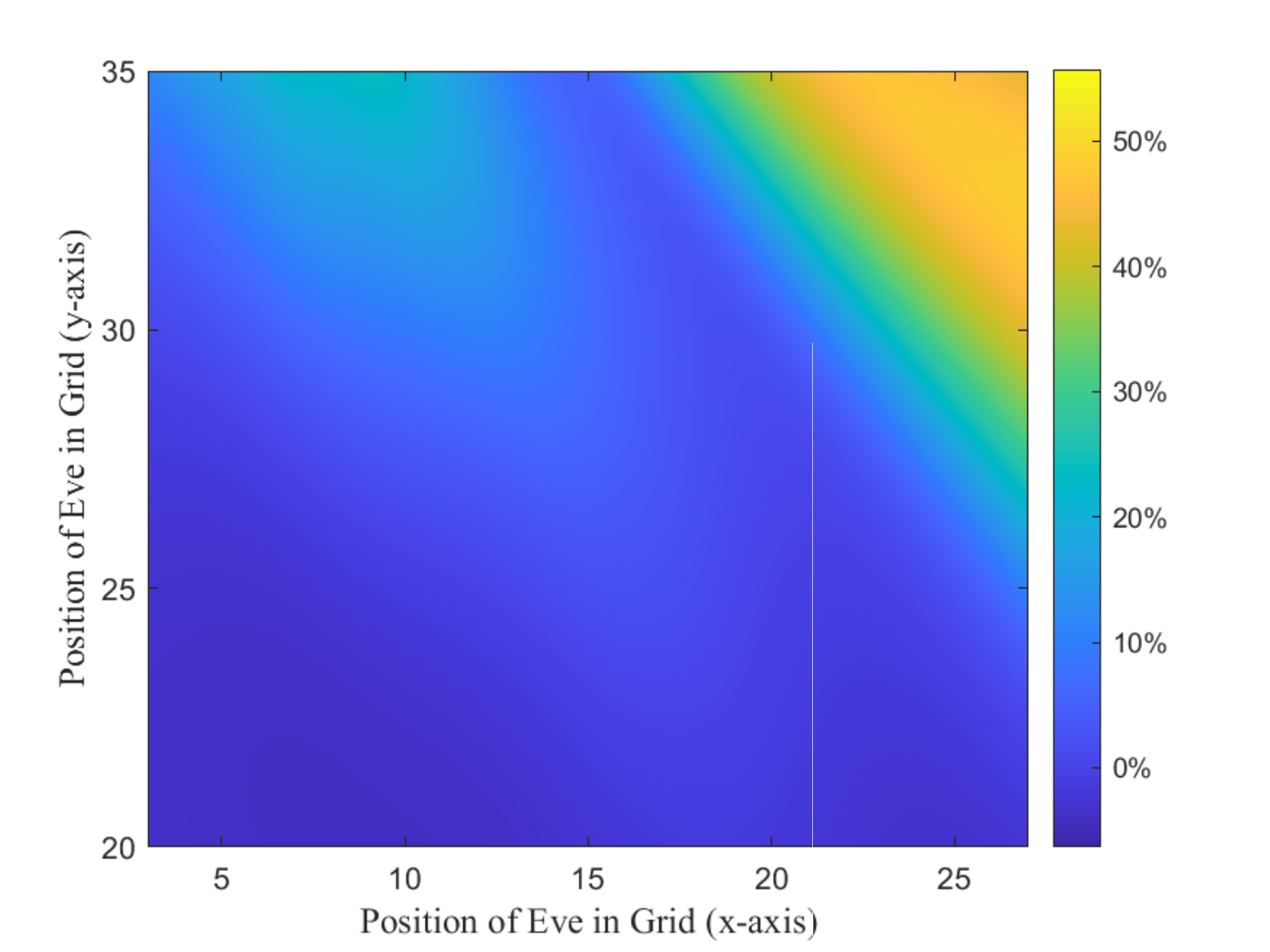}
    \caption{Spatial rate gain at Eve between the random configuration and the proposed secrecy spatial rate maximization scheme.}
    \label{fig:Heat_E_Gain_Full}
\end{subfigure}\quad
\caption{Gain in percentage in the achievable spatial rates at the legitimate RX and Eve between the random BS/RIS beamforming configuration and the proposed optimized schemes for the transmit power level $P_T= 35$ dBm.}
\label{fig:Heatmaps}
\end{figure*}

\subsection{Spatial Rate Performance Gain} \label{sec:Spatial_Gains}
In Fig.~\ref{fig:Heatmaps}, we study the impact of the RIS at the entire $\mathcal{S}_i$ regions, by evaluating the gain of the two optimized schemes with respect to the random setting scheme, as described in Section \ref{sec:Sims_Setup}. Specifically, by considering $P_T = 35$ dBm, the gains between the ``Random'' and the ``RX Only'' schemes in the achievable rates at the legitimate RX and Eve are illustrated in Figs.~\ref{fig:Heat_RX_Gain_RX_ONLY} and \ref{fig:Heat_E_Gain_RX_ONLY}, respectively, while Figs.~\ref{fig:Heat_RX_Gain_Full} and \ref{fig:Heat_E_Gain_Full} depict the respective rate gains between the ``Random'' and proposed schemes. It can be observed from all subfigures that there exist regions where the RIS enhances or degrades the rate at both receiving nodes. In addition, it is shown that the rate gains via both optimized schemes are more pronounced at the RX rather than at Eve; this is evident from the more yellow-colored areas indicating larger gains with the optimized schemes. Especially for the case of optimizing the proposed objective, the achievable gains are larger, yielding about $10\%$ larger gains between RX and Eve, as observed from Figs.~\ref{fig:Heat_RX_Gain_Full} and \ref{fig:Heat_E_Gain_Full}. On the other hand, the corresponding gains when optimizing the RIS with respect to ``RX only," span the same range of improvement, exhibiting a minimum gain below $14\%$ and maximum gain slightly above $26\%$. Finally, it is inferred from Fig.~\ref{fig:Heat_E_Gain_Full} that the largest part of $S_{\rm E}$ is degraded, apart from the sub-region where there exists a strong reflective beam. This behavior is justified by the fact that this sub-region is the most distant sub-area from the RIS, according to the described placement. It, hence, yields smaller performance gains in terms of degrading Eve's performance.

\section{Conclusion} \label{Sec:Conclusion}
In this paper, we studied RIS-enabled MISO communication systems, where a passive reflective RIS was deployed to extend the coverage of a legitimate link. We proposed a novel physical-layer security metric, which characterizes the spatially averaged secrecy spectral efficiency over a targeted geographical area where both the legitimate receiver and a potential eavesdropper can be located. Focusing on maximizing this metric, we investigated the joint optimization of the BS's linear precoder and the RIS's reflective beamformer. It was demonstrated that the proposed solution yields a fixed joint beamforming design for the entire area under investigation, which is capable of safeguarding MISO communications even for cases where the eavesdropper experiences similar pathloss attenuation with the legitimate receiver.

\section*{Acknowledgment}
This work has been supported by the EU H2020 RISE-6G project under grant number 101017011.

\bibliographystyle{IEEEtran}
\bibliography{references}

\end{document}